# The Role of Text in Visualizations: How Annotations Shape Perceptions of Bias and Influence Predictions

Chase Stokes, Cindy Xiong Bearfield, and Marti A. Hearst

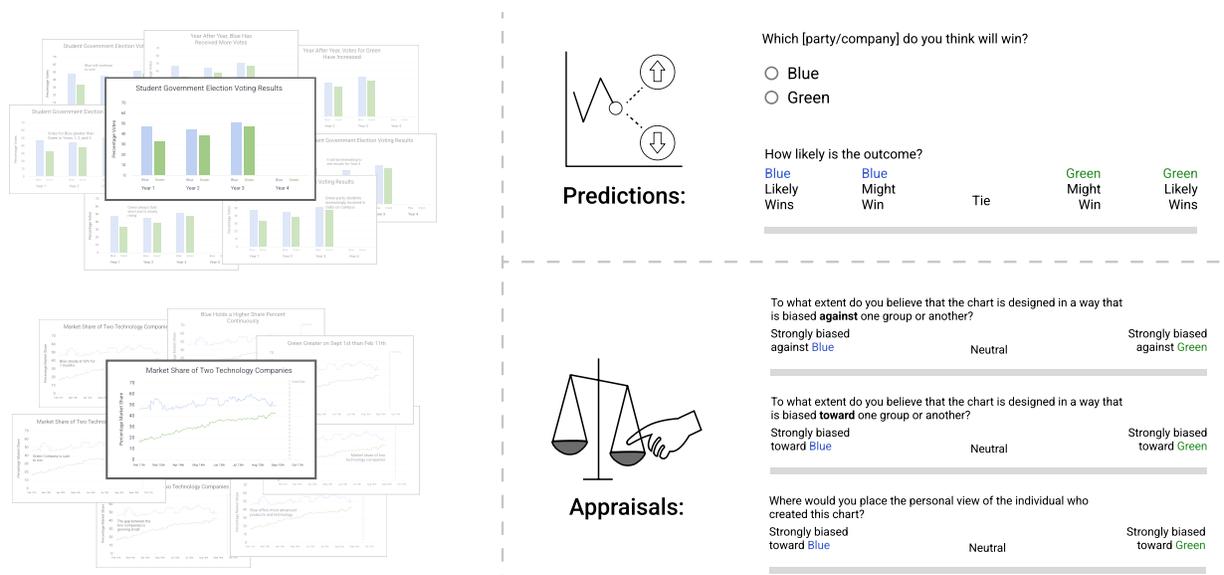

Fig. 1. (Left) Study stimuli consisted of line and bar charts that were derived from prior work and designed to have ambiguous prediction outcomes. The experiments varied the text position and text content for these charts; examples of these stimuli from both studies are shown behind the baseline charts. (Right) Two tasks were studied with crowdsourced participants: prediction of the outcome of the trend, and assessment of the bias of the visualization author using the assessment questions shown.

**Abstract**— This paper investigates the role of text in visualizations, specifically the impact of text position, semantic content, and biased wording. Two empirical studies were conducted based on two tasks (predicting data trends and appraising bias) using two visualization types (bar and line charts). While the addition of text had a minimal effect on how people perceive data trends, there was a significant impact on how biased they perceive the authors to be. This finding revealed a relationship between the degree of bias in textual information and the perception of the authors' bias. Exploratory analyses support an interaction between a person's prediction and the degree of bias they perceived. This paper also develops a crowdsourced method for creating chart annotations that range from neutral to highly biased. This research highlights the need for designers to mitigate potential polarization of readers' opinions based on how authors' ideas are expressed.

**Index Terms**—Visualization, text, annotation, perceived bias, judgment, prediction

✦

## 1 INTRODUCTION

Visualizations leverage the power of our visual system to enhance data communication; however, designing a visualization involves more than selection the best visual representation. *Textual* information, such as titles, annotations, and captions, often accompanies visualizations. Past work has shown that design choices for textual information, such as its content and position, can interact with the visual elements to influence what readers take away from the visualization [8, 20, 24, 25, 39]. However, much remains under-explored about the role of text in visualizations, leaving open a question for designers about how to convey information clearly and honestly via textual choices. Research comparing the effects of text annotations in varying contexts is needed to shed light on these choices.

Prior work has investigated the impact of design choices by text position (such as title, caption, or data annotation) [39], semantic level (from describing the axes to including real-world context) [28], and biased or mis-aligned text [24, 25]. In this work, we conducted two pre-registered empirical studies that varied text along each of these dimensions to further investigate the role of text in reader interpretations and impressions. The studies examined the influence of text on reader interpretation of visualized data for two primary tasks: participants' **predictions** about trends in the data and participants' **appraisals** of the bias of the author of the visualization. Both tasks are associated with higher-level synthesis and evaluation of information [9], which are critical to data interpretation but often understudied.

We use a visualization depicting a competition between two groups (Blue and Green) designed to be inherently ambiguous in terms of outcomes, as shown in Figure 1. For the **prediction** task, we were most interested in how participants' predictions varied if the text referred to

• Chase Stokes is with the University of California Berkeley. E-mail: cstokes@ischool.berkeley.edu.
• Cindy Xiong Bearfield is with the Georgia Institute of Technology.
• Marti A. Hearst is with the University of California Berkeley.

Table 1. Overall, we found that text has a small effect on predictions made from visualizations but a large effect on perceived bias of these visualizations. Further investigation uncovered that predictions which were unaligned with the chart led to a greater degree of perceived bias. We replicate select findings through two studies which used different text annotations. Rows colored blue indicate that the hypothesis received partial or full support.

| Hypothesis | Support | | Finding Summary |
|---|---|---|---|
| **RQ1: Does textual annotation on a chart influence predictions and appraisals in a visualization?** | | | |
| H1.1: Participants more frequently make predictions aligned with the chart than unaligned with the chart. | Study 1: Partially Supported | Study 2: Partially Supported | Participants more frequently made aligned predictions in half of the cases, with a small effect size in those cases. |
| H1.2: Participants rate predictions as more likely when aligned with the perspective in the chart. | Study 1: Partially Supported | Study 2: Partially Supported | Participants rated aligned predictions as more likely about half the time, with a small effect size in those cases. |
| H1.3: Readers more frequently make appraisals matching the chart than not matching the chart. | Study 1: Supported | Study 2: Supported | Participants consistently more frequently made appraisals which matched the perspective in the chart. |
| H1.4: Participants rate appraisals as more likely when matched with the perspective in the chart. | Study 1: Supported | Study 2: Supported | Participants consistently rated the appraisals as more likely when they matched the perspective in the chart. |
| **RQ2: How does the semantic content of textual annotations impact predictions and appraisals?** | | | |
| H2.1: External information has greater influence on predictions than other types of text content. | Study 1: Not Supported | | The content of the text did not have an effect on predictions. |
| H2.2 External information has greater influence on appraisals than other types of text content. | Study 1: Partially Supported | | External context resulted in greater rates of matched appraisals, but no other content had an effect. |
| **RQ3: How does the position of text on the chart impact predictions and appraisals?** | | | |
| H3.1: Titles have greater influence on predictions than annotations. | Study 1: Not Supported | | The position of the text did not have an effect on predictions. |
| H3.2: Titles have greater influence on appraisals than annotations. | Study 1: Not Supported | | The position of the text did not have an effect on perceptions of bias. |
| **RQ4: How does bias in text affect predictions from the data?** | | | |
| H4.1: Aligned participants rate predictions as more likely when viewing highly biased annotations in comparison to less biased annotations. | Study 2: Not Supported | | The degree of bias in an annotation did not affect readers' perceived likelihood when they were aligned with the chart. |
| H4.2: Unaligned participants rate predictions as more likely when viewing highly biased annotations in comparison to less biased annotations. | Study 2: Not Supported | | The degree of bias in an annotation did not affect readers' perceived likelihood when they were not aligned with the chart. |
| **Exploratory Analysis** | | | |
| Participants whose predictions are unaligned with the chart appraise charts as more biased than those who are aligned. | Study 1: Supported | Study 2: Supported | Participants who made unaligned predictions consistently gave higher bias appraisals on average. |

or favored one party over another. For instance, a participant looking at the bar chart must decide if Blue or Green will win the election based only on their visual perception of the chart and a neutrally worded title. Their decision rests on how they weigh two conflicting trends: for both charts Blue is consistently higher than Green, but Green is steadily increasing over time. Prior work has shown close to a 50-50 split for this determination based on the visual appearance of the charts [47].

We use these visualizations as a probe into the influence of text within visualizations, to investigate under what conditions introducing text swayed people's perception of these charts. We introduced text into two **positions**: as the title or as an annotation located next to the data. We also varied the **content** of the text. In each condition, we asked participants to indicate which side they think will prevail, and compared this prediction to what the text was implying might be the right answer. We use the terms **aligned** and **unaligned** to express the two possible outcomes. For instance, if the participant predicted Blue would prevail when the text also mentioned or favored Blue, then the visualization and the participant were aligned. If the participant predicted Blue, but the text favored Green, then they were unaligned. If the participant viewed neutral text, they were neither aligned nor unaligned with the text regardless of their response (as no outcome was supported by the text). In these studies, we found little impact of text on participant predictions.

For the **appraisal** task, the participants were asked to indicate if they thought the author of the visualization favored one side or the other (referring to the entire visualization, including but not limited to the text). If the appraisal was the same as the ground truth bias (the side favored by the text), we say the appraisal **matched** the text; otherwise it was **not matched**. For example, if the text read 'Year after year, Blue has received more votes,' the ground truth bias would be in support of the Blue party. Responses to neutral text were neither matched nor not matched (as no ground truth bias was present in the text).

This paper defines **bias** as "language favoring one side or idea over another without sufficient justification". In order to vary the text along this bias dimension, we developed and evaluated with two different sets of text stimuli. The first set (used in Study 1) was based on semantic levels as proposed by Lundgard & Satyanarayan [28] which range from describing specific chart components to bringing in context external to the chart. In Study 2, we were interested in determining if text worded in an explicitly biased way had a stronger effect than the semantic levels on participants' predictions and bias judgments. Since no such collection of text exists for visualization annotations, we developed a method to successfully crowdsource text ranging from neutral to highly biased expressions.

**Contributions:** We contribute two empirical studies examining the effect of spatial position and content of textual information across two visualization tasks: predicting future trends and assessing author bias, using simple bar and line charts with relatively neutral topics as case studies. Our findings indicate that while the interpretation of the chart is minimally impacted by text, the content of the annotations does lead to the perception of bias on the part of the visualization author. We also contribute a method for creating a set of text annotations on a spectrum from highly biased to neutral.

The results of the present studies can inform the design of visualization tools to combat cognitive biases in data analytics and storytelling via textual annotations. These results suggest existing toolkits should expand from their current focus on *visual* intervention techniques, such as showing interaction history, displaying alternative views, and highlighting under-explored data points [42, 43]. The results from this paper also add to existing literature with regards to data ethics [6, 12, 44] and should inspire reflections on best practices in generating guidelines for data storytelling to reduce the risk of miscommunication, spreading misconceptions, and reducing trust in scientific communication.

## 2 RELATED WORK

### 2.1 Integrating Text and Charts

Text can describe many aspects of a chart. Lundgard et al. [28] created a four-level semantic taxonomy for categorizing text for readers who are blind or have low vision (BLV). Yang et al. [49] created a taxonomy of text in spoken explanations which captures its ability to explain different components of complex charts, provide examples for how to interpret charts, describe the chart's construction, or add external context to the chart. But how should text be best incorporated into visualizations has remained under-explored.

In a visualization, visual elements interact with text to influence reader experiences. While visual elements clearly indicated key information in a visualization and helped readers recall data patterns such as trends more accurately [21], readers tended to fixate on text more often when visuals and texts are shown together [30]. Readers focused on the visualization titles and rely on them during information recall [5], although they were more likely to take away the content in accompanying text when it was positioned near the data, rather than in a title [39]. Kim et al. [21] used a trend prediction task to compare how well people recalled data depending on whether they had to first predict a trend or not and showed that text helped people recall data better than visualizations did. This might be because text provided an easier extraction of information [30], and thus readers strongly rely on textual information when extracting takeaways from a visualization, whether it be from a caption [20], a title [24], or an annotation [39].

Charts with text annotations and clear titles that 'focus' readers to specific takeaways tended to be perceived as more aesthetically pleasing, clear, and professional [1]. In support of these findings, studies have found that readers prefer a combination of text and visualizations compared to text or visuals alone [39]. However, the specific amount and content of the text matters. People preferred text that adds information, as opposed to unnecessary content [1], and more data-descriptive language and external information had the most influence on the language used in a reader's takeaway [39].

Recent guidelines for practitioners have incorporated these research findings to help disseminate best practices for text in visualizations. For example, Brath [7] presents a large design space for visualizing text. Setlur & Cogley [36] includes case studies and methods for effectively combining text and visualization. Knaflic [23] provides design guidelines for effective visualizations leveraging annotations for explanation.

### 2.2 Bias and Intervention in Visual Data Communication

Biases can enter a visualization in a number of ways. One lightweight framework for studying biases in visualization research is based on a 3-tier model of bias in low-level perception, decision-making or actions, and social judgments [10]. This framework acknowledges that biases can be introduced at various stages of the visualization process, from the initial perception of the visual data to the final interpretation and decision-making based on the visualization. For instance, biases in perception can occur when the reader's sensory processing introduces distortions, while action biases can occur when the reader's interpretation or evaluation of the data is skewed. Social biases can also influence the reader's decision-making process, often as a result of cultural beliefs or personal assumptions.

In the realm of textual information, biases in these areas can also be introduced through persuasive language and the author's intentions, which have long been important in the study of rhetoric and communication. An emerging body of work has demonstrated that readers can be easily biased by textual information when making sense of data. For example, readers' recall of key visualization takeaways can be biased by the title to even contradicts the intended message [25]. Narratives describing data patterns can drive people to see those patterns as more visually salient such that they miss other key patterns in the data [48].

However, not all textual information leads to bias in reader interpretations. For example, data facts can aid visualization interpretations and support data exploration [37]. When answering specific questions about data interpretation, titles (even exaggerated ones) did not have an effect on participant responses but can lead to less attention paid to graphical axes [26, 27].

The visualization community has yet to systematically explore specific text design choices and linguistic features and that elicit bias in visual data interpretation. Existing work has demonstrated that an empirical understanding of cognitive biases can lead to concrete intervention design guidelines. For example, people use mental schema to make sense of visualizations [34]. Empirical user studies have shown that charts can be misinterpreted when the correct interpretation is misaligned with one's mental schema, such as reading a chart with an inverted y-axis [31, 33]. The results of these studies informed the development of annotation tools that can combat misunderstanding of deceptive visualization online [15].

### 2.3 Salience and Belief-Driven Data Interpretation

A reader can extract many patterns from a visualization. Bottom-up feature saliency can significantly influence what people see in a visualization [18]. For example, the spatial arrangement of icon arrays can lead to over or underestimations of depicted percentages [45]. Data values encoded with visually salient color, shape, and size can pop out to a reader [17]. In a similar vein, textual information in visualization can increase the visual saliency of the particular data patterns by drawing reader's attention to that specific location [20].

In addition to bottom-up effects, readers can also be heavily influenced by their existing knowledge, belief, or motivation in their interpretations of data. Existing work has demonstrated multiple cases of this top-down effect. For example, prior beliefs about the correlation of two variables can drive people to under or overestimate their correlation on a scatterplot [46]. Political affiliation is another common factor that can drive differing interpretations of data [14]. When viewing the same chart depicting global temperature over time, liberals and conservatives attended to different areas which confirmed their preconceived notions regarding climate change [29]. When reading text on a visualization, readers could form expectations or beliefs about the data. This suggests that exposure to textual information may be another channel driving top-down mechanisms in data interpretation, but this process remains underexplored in current visualization research. These studies further inform the growing body of work examining these processes and their implications for visualization comprehension.

## 3 STUDY 1

Study 1 examined the influence of text on reader interpretation of visualized data in two primary contexts: participants' **predictions** about the data using the visualization and participants' **appraisals** of the bias of the author of the visualization.[1] We used both binary and scalar measures of predictions and appraisals, yielding four outcomes of interest. We additionally manipulated the semantic level and position (title or data annotation) of the text to examine whether these design choices increased the influence of text. Overall, we analyzed the likelihood of a participant's prediction to be **aligned** or **unaligned** and the likelihood of a participant's appraisal to be **matched** or **not matched** to the text.

---

[1]This study was preregistered on OSF prior to data collection: https://osf.io/dtnhm/?view_only=0c1f12eca87c474fa0f0f589f5c3c14b

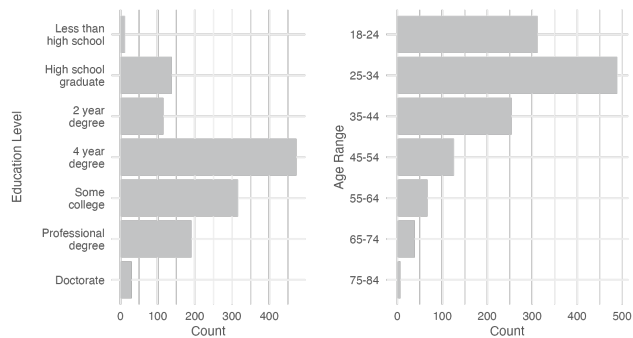

Fig. 2. Distributions of demographics for Study 1 and 2 combined. Distributions for each study followed a similar pattern as shown here.

### 3.1 Participants

To determine the proper sample size for this study, we conducted a power analysis with G*Power [16]. Means and variances in pilot data indicated a medium effect size of 0.45. Based on a power analysis (power = 0.8, d = 0.45, $\alpha$ = 0.05), the ideal sample size (post exclusions) was 640 participants (320 for bar charts, 320 for line charts).

As this is a relatively large sample size already, data quality control checks were conducted during data collection to ensure we were able to reach the ideal sample size without oversampling. This resulted in an iterative sampling process as recruiting continued until the sample after exclusion reached the desired size. In total, we recruited 653 participants on Prolific [32]. We filtered for people who were fluent in English and had an approval rate greater than 95%. 13 participants were excluded as a result of data quality control checks conducted during data collection. This resulted in the target sample of 640 participants (320 viewed bar charts, 320 viewed line charts). Participants were paid $0.75 for 3 minute study, at the equivalent of $15(USD)/hour.

Most of the participants in this study were young adults who had at least completed partial higher education. The most common age range reported was 25 - 34, followed by 18 - 24. The most common education level was a 4-year degree, followed by "some college". The full distributions are shown in Figure 2.

### 3.2 Stimuli and Design

This study used the ambiguous chart stimuli introduced in Figure 1, holding the graphical component constant and varying only the text shown on the charts. This experimental design approach allowed us to ask questions such as: If the text on the chart presents a particular interpretation of the data, are viewers more likely to report similar interpretations? If the text shown aligns with the viewer's perception of the graphics, does this correspond to an increase in their confidence in their prediction? How might the text impact viewer's assessment of the bias of visualization design and visualization authors?

We manipulated two aspects of the text: placement and content. For placement, we experimented with two common locations: **titles** and **annotations** [22, 36, 39]. Figure 3 shows the locations where text annotations were placed for each of the Blue and Green conditions in this study. Titles were placed at the top and had a font size of 1.5 times as large as annotation text.

For content, we recognize that the design of text-based stimuli can be challenging since small details in wording can have large effects on interpretation and perceptions of bias. The work by Lundgard & Satyanarayan [28] demonstrated that there exist individual differences in terms of preferred semantic levels and perceived bias in the text. For example, blind and low-vision individuals preferred text to convey information of encoding and statistical relationships, whereas sighted individuals tend to prefer text that describes perceptual and cognitive aspects with external context. Thus we constructed the text stimuli to correspond to the four semantic levels identified in Lundgard & Satyanarayan [28]:

**Semantic Level 1 (L1):** Encoded and elemental components (e.g., chart topic, axes). For example, "Student Government Election Voting Results." L1 acted as the control condition for Study 1. L1 was also used for the title for conditions where there was an annotation on the chart.

**Semantic Level 2 (L2):** Statistical and relational components (e.g., point comparisons; extrema). For example, "Votes for Blue greater than Green in Years 1, 2, and 3."

**Semantic Level 3 (L3):** Perceptual and Cognitive aspects, (e.g., trends over time, descriptions of percepts). For example, "Year after year, Blue has received more votes."

**Semantic Level 4 (L4):** External context (e.g., sociopolitical events). For example, "Blue group students highly involved in clubs on campus." This level links the text to external sociopolitical events, and is more frequently associated with potential bias than the other levels [39].

We constructed 14 sentences/phrases: two L1 phrases and four sentences each for L2-L4 (in order to represent support for Blue and Green respectively). These were places as shown in Figure 3.

### 3.3 Procedure

Participants were randomly assigned to read a chart with text, varying in three possible ways: semantic content (L1, L2, L3, or L4), position (title or annotation), and outcome supported by the text (Blue or Green). When the text was positioned as an annotation, the title was L1 content, in order to control for the influence of semantic content. When the text was positioned as a title, there was no annotation. An example of the survey taken by participants can be found in supplemental materials.

First, participants completed the **prediction** task. They predicted which of the two groups they expected to have a greater value at a given future point in the chart. The response option order was randomized and consistent throughout the survey, such that participants either saw the "Blue" option on the left or the "Green" option on the left. After this forced choice response, participants rated how likely the prediction was on a sliding scale from -25 to 25, with each side of the scale representing that one side was likely to win ("likely Blue wins" and "likely Green wins").

Next, participants completed the **appraisal** task. They rated the likelihood of author alignment ('likely Blue author" to "likely Green author") on a similar sliding scale from -25 to 25. Responses to this scale were used to construct the binary appraisal, such that positive values are coded as 'Green' and negative values as 'Blue'.

Participants also provided justification for predictions and appraisals through free-response questions, which served as a quality control check. Participant responses were excluded if the responses were inconsistent (e.g., selecting that Blue would win in the binary outcome response but that Green would win in the scale outcome response) or if they submitted nonsensical free-text responses. Finally, participants reported their demographic information including their age range (e.g., "18 - 24") and education level (e.g., "Some high school"). Responses to demographic questions are shown in Figure 2.

### 3.4 Research Questions, Hypotheses, and Results

In Study 1, we examined the three research questions in relation to predictions and appraisals, each with its own set of hypotheses. See Table 1 for a summary of the hypotheses and results. Test results are also bolded in the sections below. We summarize the major findings as they relate to our key research questions. More detailed results and testing are described in the remainder of this section and in the supplementary materials.

#### 3.4.1 RQ1: Does textual annotation on a chart influence predictions and appraisals in a visualization?

**Predictions:** Prior work from Kim, Setlur, & Agrawala has shown that when the *caption* of a chart described a particular visual or semantic feature, readers were more likely to report that feature in their *take-away* than if there were no accompanying caption [20]. We examined whether this influence of text extended to the task of making a prediction from the data. Participant **predictions were more frequently aligned than unaligned when viewing bar charts**, with 57.1% of predictions aligned with the perspective presented in the text ($\chi^2$ = 4.82, p = 0.028, Cohen's h = 0.28). However, **this effect was not true for line charts**, with only 50.6% of predictions aligned ($\chi^2$ = 0.038, p = 0.846, h = 0.03). Cohen's h [11] indicated the effect size of the $\chi^2$ testing - a small effect for the difference between bars.

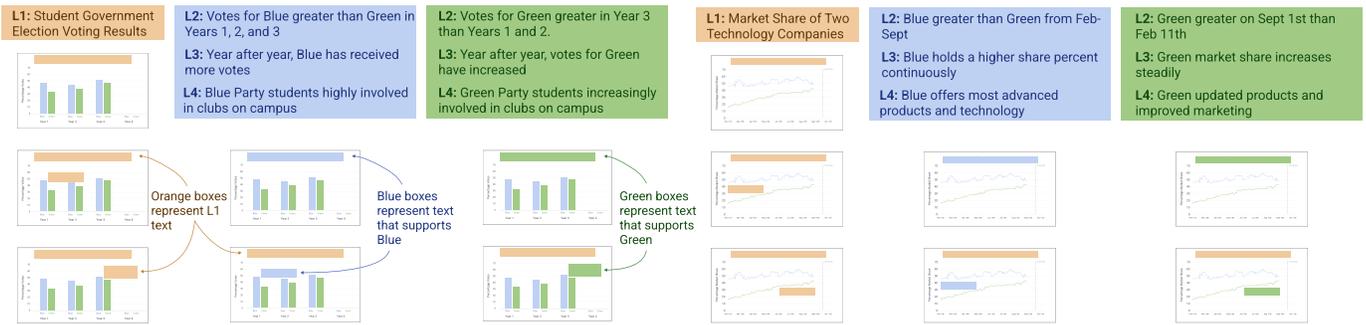

Fig. 3. Study 1 conditions for both bar and line charts. The colored rectangles on the small images of charts represent the text that was placed there. Orange indicates L1 (neutral text), blue indicates L2-L4 in support of Blue, and green indicates L3-L4 in support of Green.

Across studies on judgment and decision-making, visualizations tend to increase the decision-making confidence of participants, compared to conditions without visualizations [13]. We used the prediction likelihood ratings as a proxy for decision confidence in this study, further evaluating the effect of text on predictions. For bar charts, **prediction likelihood ratings were similar between aligned** (mean = 12.8) **and unaligned participants** (mean = 11.0). These were also similar to ratings for the control condition, which had a neutral (L1) title and no annotation (mean = 12.6). For line charts, **prediction likelihood ratings were greater for aligned participants** (mean = 13.1) than unaligned participants (mean = 11.2, p = 0.030). Neither values were different from the control condition (mean = 12.0).

These findings were evaluated with Kruskal-Wallis rank sum testing, post-hoc Dunn testing and Bonferroni correction (bar: K-W $\chi^2$ = 5.70, p = 0.058, $\eta^2$ = 0.011; line: K-W $\chi^2$ = 6.70, p = 0.035, $\eta^2$ = 0.015). $\eta^2$ values indicated a small effect size (large = 0.14, medium = 0.06).

Overall, we found mixed results when comparing our findings to those from prior literature. Specifically, text seems to have only a minor effect on predictions, both in terms of the prediction itself and the perceived likelihood of the event. This differs from text's impact on takeaways as shown in prior work [20, 39, 50] but is consistent with findings that exaggerated text has limited effect on comprehension question [26]. Additionally, there was not consistent change in likelihood ratings, likely due to the minimal effect of the text on the prediction.

**Appraisals:** In prior work examining preferences for text content in visualizations, some readers reported that text that signaled the authors' perspective or stance was either not useful [28] or disliked [38]. This indicates that readers are able to identify the author's stance from the text paired with the chart in some cases. From this, we examined whether readers' perceptions of bias were consistent with the perspective presented in the chart's text.

Participant **appraisals were more frequently matched for both chart types**, with 67.9% of appraisals matched when reading bar charts and 61.5% of appraisals matched when reading line charts. For example, when the text supported the 'Blue' group, participants more frequently perceived the visualization author as part of the 'Blue' group. (Chi-square testing: bar: $\chi^2$ = 30.82, p < 0.001, h = 0.73; line: $\chi^2$ = 12.66, p < 0.001, h = 0.46). Values of Cohen's h indicated a large effect size for the bar results and a moderate effect size for the line results.

For **both chart types, appraisal likelihood ratings were higher when participants matched** (bar: mean = 8.24, line: mean = 8.24) than when unmatched (bar: mean = 3.08, p < 0.001; line: mean = 3.42, p < 0.001). Matched appraisals also had higher likelihood ratings than control (bar: mean = 3.15, p < 0.001; line: mean = 4.10, p < 0.001). Overall, matching the appraisal to the chart perspective resulted in an increased perceived likelihood of the appraisal.

These findings were evaluated with Kruskal-Wallis rank sum testing with post-hoc Dunn testing and Bonferroni correction (bar: K-W $\chi^2$ = 51.14, p < 0.001, $\eta^2$ = 0.16; line: K-W $\chi^2$ = 29.34, p < 0.001, $\eta^2$ = 0.09). The $\eta^2$ values indicated a large effect size for the bar results and a moderate effect size for the line results.

Overall, these results are consistent with the implications from prior work, as participants consistently and confidently match bias present in the chart with the perspective presented in the text.

### 3.4.2 RQ2: How does the semantic content of textual annotations impact predictions and appraisals?

We varied the semantic content of the textual annotations following the framework proposed by Lundgard & Satyanarayan [28] as described in Section 3.2. Based on these variations in text content, we proposed the following two hypotheses:

**Predictions:** In prior work examining the influence of text on take-aways from charts, including external context (L4) (e.g., sociopolitical information) in the text (titles, annotations, and captions) had the greatest impact on the content of the takeaway [20, 39]. We investigated the effect of different semantic levels on prediction.

This analysis used a stepwise regression approach with mixed-effect models, which can be found in supplementary materials. These models predicted the likelihood ratings made by participants and incorporated random effects of the response option order and the group supported by the text. Models were compared with an ANOVA, and pre-registered exploratory models were included in this process. Including exploratory models allowed us to use the most optimal model for all analyses.

The optimal model, which was pre-registered as exploratory, only included the chart type and the alignment of the prediction. Increasing the complexity of the model by adding text level and position did not improve the performance. This suggested that **the level and position of the text did not have an effect on prediction likelihood ratings**. This model did not include demographics, as they also did not improve the performance of the model (p = 0.065).

Additionally, **the level of text did not have an effect on prediction alignment** (L4: 54.7% aligned, L3: 57.1%, L2: 49.7%; exploratory Chi-squared testing: $\chi^2$ = 1.820, p = 0.403). There was no effect of variations in text content on predictions overall. This is consistent with the divergence from prior work that we observed in other testing on predictions.

**Appraisals:** In a study examining preferences for types of text content in alternative text descriptions, blind and low-vision (BLV) participants disliked receiving L4 content in the descriptions [28]. Further investigation indicated that this was due to the possible bias imbued by the author in choosing the specific external context to include.

A stepwise regression analysis was completed, in this case predicting the appraisal likelihood ratings. Full results can be viewed in supplementary materials. The optimal model included the chart type, interaction between appraisal match and semantic level, interaction between appraisal match and position, the alignment of the prediction, and random effects. This model did not include demographics, as they did not improve the performance of the model (p = 0.250).

**L4 did not have a greater influence on the appraisal likelihood ratings** than L3 (p = 0.310) or L2 (p = 0.317). Participant **appraisals were frequently matched when viewing L4** (73.0%) in comparison to L3 (63.2%) and L2 (58.0%), which indicated some influence of text content on bias appraisals (exploratory Chi-squared testing: $\chi^2$ = 8.03, p = 0.018). This suggests that external context (L4) acts as a greater signal of author bias than other text content. This is somewhat consistent with the prior work that indicates external context can act as a greater signal of bias, but it is not a large effect.

#### 3.4.3 RQ3: How does the position of text on the chart impact predictions and appraisals?

We examined two common expressions of text on a chart: titles and annotations. Prior work found that titles were the most visually salient component of a visualization and were generally viewed first when looking at a visualization or infographics [5]. However, readers were generally more likely to match the text provided in their takeaways when the text was positioned near the data, rather than as a title [39]. Due to this conflict in prior work, we evaluate both sides of the hypothesis. These analyses used the same mixed-effect models as described in RQ2 to investigate predictions and appraisals.

**Predictions:** Position was not included in the optimal model, and thus titles did not have a greater influence on the scalar measure of predictions. A similar percentage of predictions were aligned when viewing titles (56.0%) as when viewing annotations (51.7%). This was tested with exploratory Chi-squared testing ($\chi^2$ = 0.740, p = 0.390). **Variations in the position of text did not have an effect on participant predictions.**

**Appraisals:** Titles did not have a greater influence on the appraisal likelihood ratings (p = 0.812). A similar percentage of appraisals matched the perspective in the chart when viewing titles (61.4%) as when viewing annotations (68.1%). This was tested with exploratory Chi-squared testing ($\chi^2$ = 2.04, p = 0.153). **Variations in the position of text did not have an effect on participant appraisals.**

### 3.5 Exploratory Analysis

We conducted additional exploratory analyses to examine the relation between predictions and bias appraisals. We were primarily interested in analyzing whether the alignment of the prediction made by the participant affected the appraisal likelihood rating. In other words, were participants attributing bias to the author in part because they disagreed with the perspective in the chart? When analyzing appraisal likelihood ratings, the prediction alignment variable was included in the optimal model, indicating that it did improve performance.

Participants responded with lower appraisal likelihood ratings if they made aligned predictions than if unaligned, by an average of 1.24 points (se = 0.60, p = 0.039). This change is about 5% of the total scale. This supports that bias appraisals depended in part on the participant's own alignment with the text presented in the chart.

### 3.6 Discussion

The results from this study presented an interesting interplay between text presented on a chart, predictions made from the chart, and appraisals of bias present in the chart. The impact of text on predictions was small and inconsistent, indicating that participants are likely basing their predictions primarily on the visual elements of the chart. Compared to existing findings that suggest participants rely more heavily on text when generating key takeaways from data [20, 24, 38], our findings suggest that the underlying mechanisms for a prediction task may be fundamentally different from a takeaway generation task.

This corroborates prior work demonstrating that text and visuals affect different types of data analytic tasks. As discussed in Section 2, Kim et al. [21] found that visuals had more impact than text on a prediction task, but text helped people recall details better. Ottley et al. [30] found that people did not integrate information from text and visualizations together across representations. Different tasks with data visualization may be informed by textual elements *and/or* visual components, underscoring the need for a comprehensive understanding of both aspects.

In Study 1, while the effect on prediction was not large, the text provided a clear signal of bias. Participants frequently appraised the bias to match the perspective in the chart, indicating an ability to identify an author's perspective with relative accuracy. Furthermore, matching the appraisals of bias led to significantly greater likelihood ratings, with moderate to strong effect sizes. Overall, the text on a visualization may not have an effect on the predictions made from the data, but it does have an effect on the bias appraisals and can signal the perspective of the chart's authors.

Prediction and appraisal tasks also seem to interact. Participants whose predictions were aligned with the presented text appeared to appraise the author as less biased than those whose predictions were unaligned. This potential insight corroborates existing findings from social psychology showing that people tend to agree with others who share the same opinions [41] and thus find their perspective to be less biased [35].

## 4 STUDY 2

In Study 1, we examined the effect of text on predictions and bias appraisals. In Study 2, we manipulated bias as a variable, using a new set of text stimuli ranging from highly biased to neutral text. Unlike Study 1 which varied content according to semantic levels, Study 2 focuses specifically on effect of biased text. Using these new text stimuli, we revisit RQ1 to verify consistency across text stimuli. We also investigate a new research question RQ4: *How does bias in text affect predictions from the data?*

### 4.1 Participants

Based on the effect sizes from Study 1, we conducted a power analysis to determine the proper sample size for Study 2b. The proper sample size to detect the effect is 660 participants (330 per chart type).

Responses were filtered and excluded following the same protocols as in Study 1. This resulted in 35 participants being excluded from those who viewed the bar chart and 23 participants from the line chart. As in Study 1, data quality checks were completed during data collection to avoid oversampling. We iteratively sampled until we meet the 660 participant (330 bar, 330 line) requirement post-exclusions. Participants were also paid at similarly to Study 1: $1.00 for 4 min study, a rate of consistent with the current minimum wage rate at $15 per hour.

Most participants in this study were also young adults who had completed at least some college. This is also reflective of the common demographics on the Prolific platform. The most common age response was "25 - 34". The most common education level was a 4-year degree. Distributions of participant demographics can be seen in Figure 2.

### 4.2 Stimuli

We controlled for the perceived level of bias of the text by developing a new set of text stimuli based on annotations written and assessed by crowdworkers, explicitly designed to range from neutral to highly biased. Figure 4 shows the produced texts, ordered according to their perceived bias for both the bar chart (left) and the line chart (right). We asked crowdworkers to write neutral, low-bias, and high-bias text from each of the Blue and the Green perspectives. The figure shows that the construction successfully created a spectrum of bias appraisals, as judged by a second set of crowdworkers (those sentences appraised as having higher bias appear at the top).

This construction also situates the wording of the text sentences from Study 1, showing that they fall into a range of perceived bias, with L2 perceived as low bias and L4 perceived as high bias (shown in purple in Figure 4).

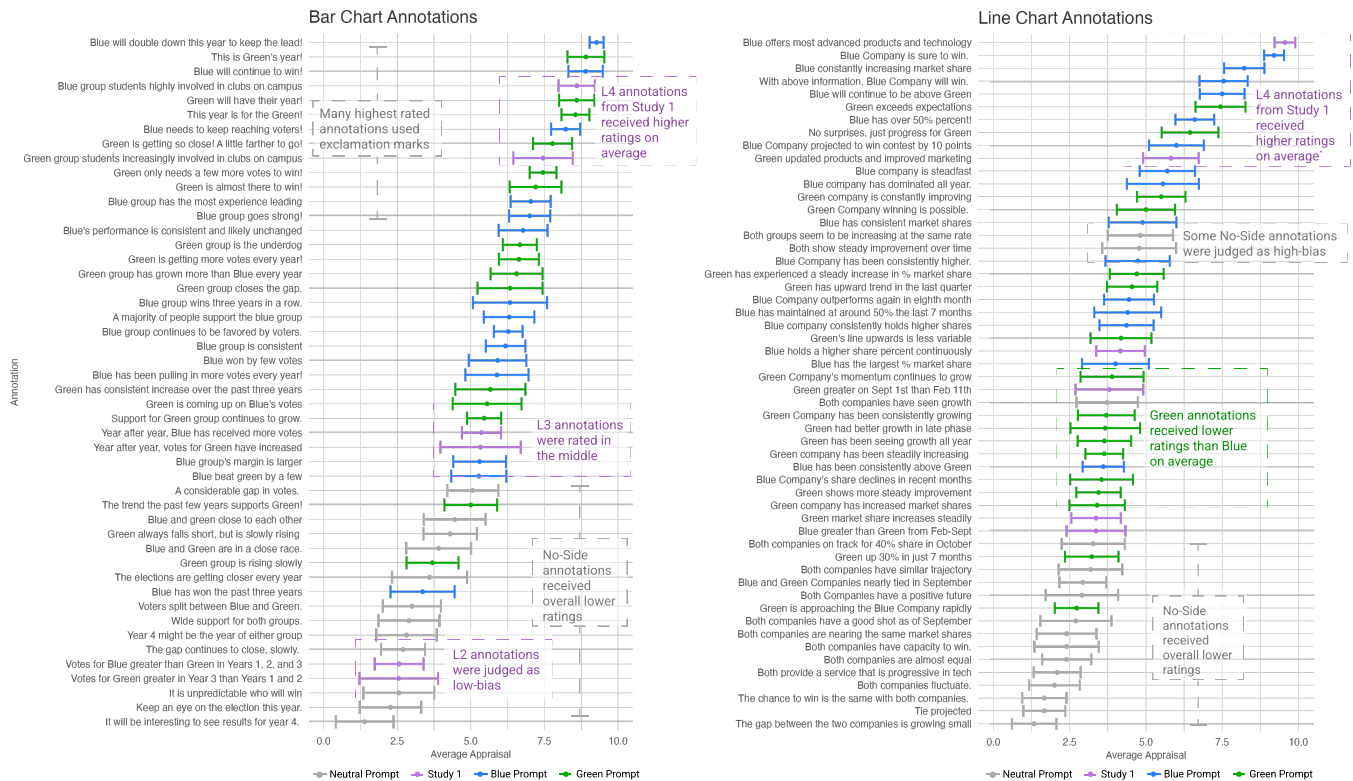

Fig. 4. Average ratings of crowdsourced annotations on a scale from 0-10 in response to the question "To what extent does the annotation favor one side without sufficient justification?" One set of crowdworkers wrote the annotations and another, separate set rated them. Ratings covered the range of values provided. Error bars show standard error on either side of the mean. Color coding indicates if the text was written from a neutral (grey) perspective, from the perspective of Blue or Green, or if the annotation is from Study 1 (purple).

We recruited 20 participants on Prolific with a background in English Language, English Literature, Communications, or Education. We chose crowdworkers with these backgrounds based on prior work that suggests than expert writers are better than non-experts at judging the quality of creative work [3, 4].

Participants viewed charts with an L1 title from Study 1 and a single gray box indicating where the annotation would go. They were prompted with the following instructions: "Regardless of your personal opinion, imagine you are a publicist working for the [Blue Group, Green Group]. Help them draft possible annotations to add to this chart that supports the [Blue Group, Green Group] winning. The annotation will be placed in the box indicated on the chart and thus has a character limit." Neutral annotations were from the perspective of a Neutral Organization but used similar instructions.

Participants were asked to write 6 annotations each: 2 from each group's perspective. Low-effort annotations were identified as those that did not mention the relevant parties for the Blue and Green questions as well as those that did not refer at all to the topic of the chart (e.g., "ours for the taking"). We also excluded annotations that contained incorrect information (e.g., "Both groups seem to be increasing at the same rate"). After exclusions, this resulted in a collection of 102 annotations including the original 12 texts from Study 1 (48 for bar chart, 54 for line chart).

To rate the bias of these 102 annotations, an additional 37 participants from Prolific.co [32] appraised annotations on a scale from 0 to 10 in response to the question "To what extent does the annotation favor one side without sufficient justification?" Four annotations were re-rated at the end of the survey as quality control. Each participant appraised over 30 annotations (32 for bars, 36 for lines). On average, each annotation received roughly 10 appraisals (bar = 10.42, line = 10.37).

This approach to stimuli creation has been employed similarly in prior work [24] and provides a range of benefits. Crowdsourcing annotations can bring more diverse viewpoints and interpretations, reflecting a wider range of understanding and perspectives than a single individual or a small team might offer. As such, they can introduce novel, creative ways of describing and interpreting data that the original authors might not have considered. This can lead to more innovative approaches to data annotation and visualization. More details about the method can be found in the supplementary materials.

As can be seen in Figure 4, overall, the spread of appraisals demonstrates a distinction between annotations with relatively high bias and those with relatively low bias. Text written from the perspective of a neutral party received lower (rated less biased) appraisals (mean = 2.97) than those written from a Blue or Green party perspective (mean = 5.77). Annotations for Study 1 received a relatively wide range of appraisals, which stemmed from the differences between the semantic levels (L2 = 3.07, L3 = 4.56, L4 = 7.86).

### 4.3 Procedure

Figure 5 shows how selected sentences were placed on the chart stimuli for Study 2. To create the text stimuli for Study 2, we first manually coded the texts that did not mention either the Blue or Green or mentioned both as having an equivalent chance to win with a no-side code. Then, we selected the four with the lowest appraisals to create the No-Side condition.

From the annotations which did *not* receive the no-side code, we selected the four with the lowest average appraisals to create the Low-Bias condition and the four with the highest average appraisals to create the High-Bias condition. Because these were written from the perspective of either Green or Blue, we augmented the crowdworker-created set with additional sentences annotations to represent the opposing side for each annotation. For example, as shown in Figure 5, one of the High-Bias annotations read: "Blue will double down this year to *keep* the lead!" The opposing side, edited manually, was "Green will double down this year to *take* the lead!" (emphasis added).

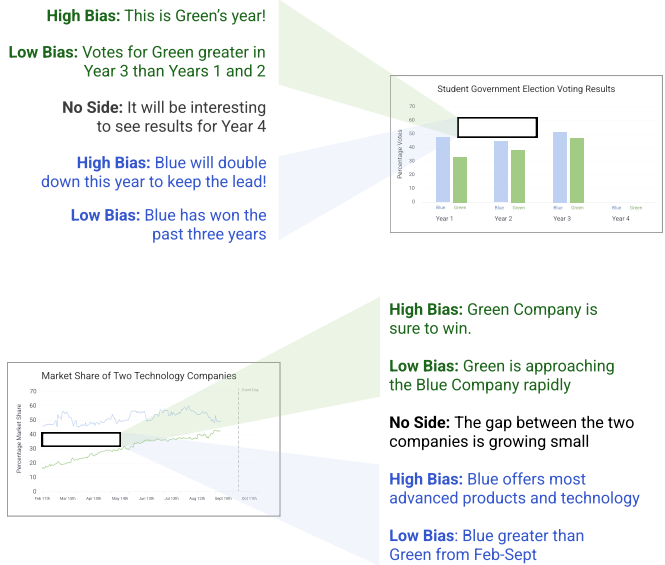

Fig. 5. Depiction of Study 2 annotation position and content for both chart types. This image depicts example annotations, as there were four possible annotations for each one shown here. Titles were always neutral in this study.

Participants completed a survey with three main sections. First, they consented to the survey and were introduced to the chart. Next, they were randomly assigned to read a chart with text, varying in two ways: degree of bias (No-Side, Low-Bias, High-Bias), and outcome supported by the text (Blue, Green). An example of the survey taken by participants can be found in supplemental materials. This text was always positioned as an annotation. Due to a lack of effect found in Study 1, we did not manipulate text position in Study 2. Examples of these annotations are shown in Figure 5. The title of the charts was always the L1 content from Study 1 (e.g., "Student Government Election Voting Results").

After viewing the chart, participants reported their predictions and appraisals in a randomized order. Prediction measures were the same as Study 1, with the addition of the question, "What is the percent chance of each possible outcome occurring?" Participants entered a value for the Blue group winning, the Green group winning, and a tie. These values were required to sum to 100.

When making appraisals, participants responded to three questions. This was a modification of the procedure in Study 1. The first two asked, "To what extent do you believe that the chart is designed in a way that is biased [against, toward] one group or another?" The third question was a modification of the appraisal from Study 1: "Where would you place the personal view of the individual who created this chart?" In the final section of the survey, participants entered their age range and education level.

### 4.4 Research Questions Hypotheses, and Results

We examined[2] two research questions in Study 2, including a repeated evaluation of RQ1 from Study 1. A summary of results can be found in Table 1, with specific test results bolded in the section below. Further details on testing can be found in supplementary materials.

#### 4.4.1 RQ1: Does textual annotation on a chart influence predictions and bias appraisals in a visualization?

**Predictions:** As in Study 1, we observed inconsistent findings between bar and line charts, but in the opposite direction. Participant **predictions**

---
[2]This study was preregistered on OSF prior to data collection: https://osf.io/detp8?view_only=e3f026a9082c413fbd5504b356ff0f66

**were more frequently aligned than unaligned when viewing line charts**, with 58.2% of predictions aligned ($\chi^2 = 5.89$, p = 0.015, h = 0.33). However, **this effect was not true for bar charts**, with only 56.0% of predictions aligned with the chart ($\chi^2 = 3.07$, p = 0.080, Cohen's h = 0.24). In Study 1, the effect was significant for bar charts but not for line charts. These inconsistent results, coupled with the values of Cohen's h indicated relatively low effect sizes for both lines and bars.

We evaluated prediction likelihood ratings in two ways: the likelihood of the chosen outcome occurring, and the likelihood each possible outcome occurring. Testing found **indication of differences for both chart types** (bar: K-W $\chi^2 = 6.25$, p = 0.044, $\eta^2 = 0.01$; line: K-W $\chi^2 = 6.28$, p = 0.043, $\eta^2 = 0.01$). **For line charts, aligned predictions had higher likelihood ratings** (mean = 13.7) **compared to control** for the original prediction question (No-Side; mean = 11.7, p = 0.039). **For bar charts, aligned predictions had higher likelihood ratings compared to unaligned predictions** for the added prediction question (p = 0.029).

Overall, these findings are consistent with a minor effect size (low $\eta^2$ values) and limited impact of text on prediction outcomes, aligning with the conclusions from our first study. These results also further support findings which indicate that both exaggerated and control titles have little effect on questions regarding the extent of an effect shown in a visualization [26].

Responses to the original likelihood question were evaluated with Kruskal-Wallis rank sum testing, post-hoc Dunn testing, and Bonferroni correction. Responses to the additional likelihood questions were evaluated with Mann-Whitney U-tests, as preregistered. The difference in testing strategy comes from the focus on specific pairwise comparisons for the added likelihood questions.

**Appraisals:** Participant **appraisals were more frequently matched for both chart types**, with 70.0% of appraisals matched when viewing bar charts and 71.4% of appraisals matched when viewing line charts. This supported the findings from Study 1. This **effect was also greater for High-Bias annotations in comparison to Low-Bias annotations**, demonstrating that participants were able to more clearly match the appraisal with the perspective when the bias was greater. (bar: $\chi^2 = 35.2$, p < 0.001, h = 0.82; line: $\chi^2 = 40.16$, p < 0.001, h = 0.88). These values of Cohen's h indicated large effect sizes.

When evaluating this hypothesis, the response to the biased "against" question was reversed, and responses to all appraisal questions were averaged for the final appraisal rating. For **both chart types, participants reported greater appraisal likelihoods when matched** (bar: mean = 7.54, line: mean = 7.51) **than when unmatched** (bar: mean = 1.77, p < 0.001; line: mean = 3.03, p < 0.001). **Matched appraisals also had a higher likelihood than control** (bar: mean = 2.78, p < 0.001, line: mean = 2.99, p < 0.001). Overall, matching the appraisal to the chart perspective resulted in an increased perceived likelihood of the appraisal, as observed in Study 1.

These findings were evaluated with Kruskal-Wallis rank sum testing with post-hoc Dunn testing and Bonferroni correction (K-W $\chi^2 = 63.38$, p < 0.001, $\eta^2 = 0.19$; line: K-W $\chi^2 = 51.2$, p < 0.001, $\eta = 0.15$). These $\eta^2$ values indicated a large effect size. Overall, these results were consistent with Study 1 and the finding that participants reliably and confidently correlate the bias in the chart to the perspective displayed in the text.

#### 4.4.2 RQ4: How does bias in text affect predictions from the data?

Study 2 expands the formal consideration of bias in text. Exploratory analysis in Study 1 indicated a possible interaction between a reader's bias appraisal and their prediction. As such, we manipulated the bias present in the text in Study 2, displaying three conditions of bias: No-Side, Low Bias, and High Bias. Here, we explored how these conditions affected predictions, as displayed in Figure 6.

Prior work from Taber & Lodge indicates that arguments aligned with the participant's own attitudes are interpreted as stronger than arguments that are not aligned [40]. In previous work on crowd-sourced fact-checking from Allen, Martel, & Rand, fact-checkers were more likely to correct misinformation from an opposing political party [2]. Accordingly, we anticipated an increase in prediction likelihood for both aligned participants when viewing High-Bias annotations. Aligned participants may see these arguments as stronger due to the more explicit support of their own attitudes, while unaligned participants may discount the text due to its higher degree of bias.

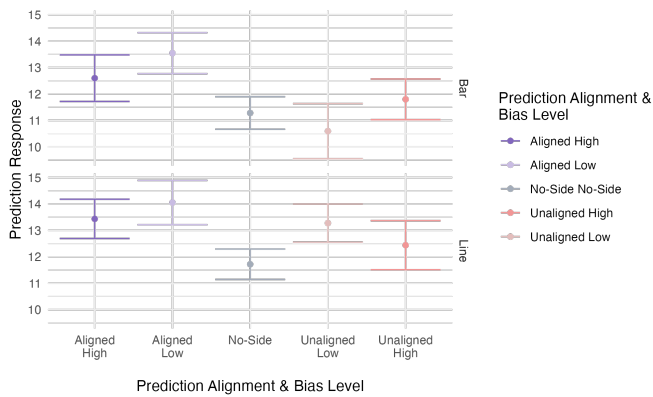

Fig. 6. Prediction confidence ratings between bias conditions and alignment. Error bars show standard error. Overall, there was an increase in responses that were aligned, but no clear differences between High- and Low-Bias conditions.

Overall, we did not find support for these predictions, as High-Bias annotations had a relatively minimal effect on prediction likelihood ratings. Average values and standard errors across conditions can be seen in Figure 6.

**For both chart types, aligned participants reported similar prediction likelihoods when viewing High-Bias conditions in comparison to Low-Bias or No-Side conditions.** However, **aligned participants reported higher prediction likelihoods when viewing Low-Bias annotations in comparison to No-Side conditions** (bar: p = 0.078, line: p = 0.048). Although the results for the bar chart did not reach significance, and the line chart only just, the trend present and the values displayed on the left side of Figure 6 indicate that, for aligned participants, Low-Bias annotations tend to have an effect relative to No-Side annotations. High-Bias annotations, on the other hand, did not have this effect in comparison to No-Side annotations. In other words, the high degree of bias present in annotations seems to reduce the already minimal impact of annotations on predictions.

These findings were evaluated with Kruskal-Wallis rank sum testing with post-hoc Dunn testing and Bonferroni correction (bar: K-W $\chi^2$ = 5.11, p = 0.078, $\eta^2$ = 0.01; line: K-W $\chi^2$ = 6.53, p = 0.038, $\eta^2$ = 0.02). The $\eta^2$ values indicated a small effect size.

**Unaligned participants reported similar prediction likelihoods across the three bias conditions** (bar: K-W $\chi^2$ = 0.642, p = 0.726, $\eta^2$ = -0.01; line: K-W $\chi^2$ = 2.42, p = 0.298, $\eta^2$ = 0.002). These $\eta^2$ values indicate a very small, negligible effect size. Average values and spread for unaligned participants can be seen on the right side of Figure 6.

### 4.5 Exploratory Analysis

As in Study 1, **participant reported lower appraisal likelihood ratings if their prediction was aligned than if it was unaligned**, by 2.03 points (se = 0.56, p < 0.001). This difference is about 8% of the total scale. This analysis was performed using the same model structure as the exploratory analysis in Study 1. These results support the idea that the appraisals of bias may vary depending on the participant's own alignment with the perspective of the chart.

Additionally, the bias manipulations did have an effect on the appraisals overall. **The appraisal match frequency was greater for High-Bias** (81.8%) **than Low-Bias annotations** (59.5%). Additionally, **appraisal confidence was higher for High-Bias matched appraisals than Low-Bias**, as can be seen in Figure 7. This further supports the ability of readers to extract information about author bias from the text presented with the chart.

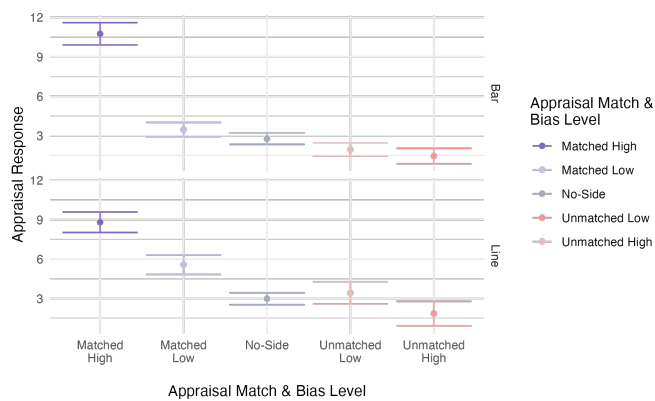

Fig. 7. Appraisal confidence ratings between bias conditions and match. Error bars show standard error. Overall, participants were not confident in appraisals which did not match the text shown. When the appraisals were matched, participants had much more confidence when appraising High-Bias text than Low-Bias or No-Side.

Bar chart annotations used language with more exclamation marks and often seemed to be written for a more casual audience. Beyond this, there were no clearly observable differences between the annotations; future work and exploration should consider linguistic differences between High-Bias and Low-Bias annotations.

### 4.6 Discussion

This study aimed to further illustrate the interplay between data predictions and bias appraisals. Many takeaways from Study 1 were supported, including exploratory analyses that unaligned participants seemed to report greater appraisals. The participant's own outcome predictions seemed to influence how biased they deemed the visualization to be, a process similar to polarization [19].

This study also corroborated findings from Study 1 that text had a small effect on predictions. Low-Bias annotations have more influence than No-Side annotations on predictions; increasing bias from Low-Bias to High-Bias seemed to reduce this effect. The limited effect of text points to the possibility of a task-dependent interaction of text and visualization. Some tasks (e.g., takeaways, bias appraisals) may rely more on textual components than visual components, while others tasks (e.g., predictions) may rely more on visual elements.

While the effect on predictions was small, the effect on appraisals was large. Participants frequently matched their appraisal with the perspective in the chart. They also reported higher appraisal likelihood ratings when they were matched with the chart. Appraisals were greater for High-Bias annotations, indicating that readers differentiate between Low- and High-Bias content.

## 5 DISCUSSION, LIMITATIONS, AND FUTURE WORK

We examined the effect of the position and content of textual information across two visualization tasks using simple bar and line charts: a prediction task where participants judged the future state of data and an appraisal task where they assessed the degree of author bias. We controlled for the effect of prior beliefs by using neutral topics and completed thorough analyses to determine the nuanced effect of text on these tasks.

The results of both Study 1 and Study 2 highlight the nuanced roles of text and visuals in chart interpretation and have significant implications in areas like data journalism, education, and business analytics. While visual elements guide predictive analyses, textual content shapes perceptions of bias. This emphasizes the need for professionals to carefully balance text and visuals to convey accurate, unbiased information.

Journalists should also consider how text influences bias perception in efforts to guide balanced storytelling. Educators can leverage these insights to teach more effective data interpretation skills, focusing on how different components of a visualization contribute to understanding.

This research also emphasizes potential key choices for developers and designers of data visualization tools. It seems important to equip visualization and visual analytics tools with features that enable users to customize annotations in a manner that aims to reduce bias while effectively conveying intended messages. However, there are several limitations to our approach that open up opportunities for future research, which we discuss below.

**Participant Pool:** Participants in this study tended to be educated young adults on average, which may have had an influence on the decisions made with visualizations. While some of the analyses incorporated demographics and did not find a significant effect of including them, it was not possible to account for these variations in all analyses. As such, results may vary with different groups of participants, particularly those who are older and/or have a lower level of education. Future work should target the decision-making processes in these populations, as they are underrepresented in not only the studies presented here but in many visualization studies.

**Alternative Charts and Topics:** We kept a consistent topic and used relatively simple line and bar charts in our experiments. This means the driving factor behind differences in participant behaviors across both the prediction task and the appraisal task could stem from chart type *or* chart topic. In some cases, we observed text influencing reader predictions to be more frequently aligned with what the text stated.

However, this effect was not consistent across chart types or experiments. In Study 1, the effect was significant for bar charts. In Study 2, the effect was significant for line charts. We speculate that this difference is driven by the overall small effect of text on predictions, supported by effect sizes for the relevant tests.

To account for the potential effects of chart type and topic and to test for generalizability, future research should explore a wider set of chart types and chart topics. By additionally investigating how participants behave when the topic has higher real-world relevance across a wider variety of charts, we can identify the effect of text that is general versus chart- or domain-specific.

**Elicitation Methods for Prediction Tasks:** We also found differences in participants' predictions elicited by the two different questions assessing prediction: the binary choice and the likelihood rating. This is clearest in Study 1. Binary choices showed an effect for bar charts, but likelihood ratings showed an effect for line charts. This difference may be driven by the response modality. For example, the binary choice could polarize participants to report an opinion more extreme than their actual opinion. Future work should further evaluate the effect of response modalities in eliciting participant interpretations of visualizations. Researchers can identify other real-world tasks performed with data visualizations and design experiments by simulating those tasks.

**Top-Down vs. Bottom-Up Effects:** We did not design the current study to tease apart the effect of top-down belief-driven processes from bottom-up saliency-driven processes in prediction and appraisal tasks. Both processes are likely at play when readers complete prediction and appraisal with textual annotations. A reader may latch on to a particular visual component in the visualization (bottom-up) to form a belief to be compared to the textual annotation, and then make a decision based on whether the text support or contradict the belief they formed. Alternatively, a reader may rely on the textual annotation and then seek patterns that either support or contradict it (top-down), and then make their decision based on the ease of finding supportive or contradictory patterns. Future work can investigate the interaction between top-down and bottom-up effects in data interpretation with text by manipulating the visual saliency of data patterns or capturing the sequence of reasoning strategies participants make decisions.

**Expanding Text Design Space:** This paper took an initial step in exploring the interplay between text and visualization. Prior work has shown that the semantic frame can profoundly impact people's interpretation of a visualization, but individual people can adopt very subjective interpretations of a given textual frame. Future work should explore other design spaces of text in visualization, such as the visual styling of the text and the intended purpose of the text. For example, textual information intended to mitigate misunderstanding might impact interpretation in a different way compared to textual information intended for persuasion. Future examinations of text design should also consider sophisticated models to determine signals of bias present in text.

## 6 CONCLUSION

Two empirical studies find that text information seems to have a minimal effect on how people see trends in data to make predictions, but a substantial effect on how biased they perceived the author to be. These studies found that visualization readers can be quite sensitive to the degree of bias in textual annotations. Furthermore, perceived author bias appears to become more exaggerated when the content is incongruent with the participant's own perspective. These results are exploratory and should be treated as such; they offer a clear direction for future research and important implications for possible data understanding. These results add to the existing literature on the interplay of text with visualization, as well as help inform questions surrounding data ethics and storytelling.

While the findings of this study pave the way for further explorations into annotation content and structure across domains, this research also underscores the necessity for more sophisticated models to analyze annotations and possible biases. Visualization authors would benefit from an advanced model capable of quantifying bias based on word usage and data patterns. This work acts as a stepping stone to inform how those models should conceptualize bias or capture responses.

**ACKNOWLEDGMENTS**

This research was supported in part by a gift from the Allen Institute for AI, by NSF awards IIS-2237585 and IIS-2311575, and by a National Science Foundation Graduate Research Fellowship Program under Grant No. DGE 2146752. Thanks also to Simone Laszuk for survey piloting and manuscript refinement.

**Chase Stokes** is a graduate student at UC Berkeley School of Information. He studies information visualizations and text, with particular interest in improving design practices for combining visual and textual information.

**Cindy Xiong Bearfield** is an Assistant Professor in the School of Interactive Computing at Georgia Tech. By investigating how humans perceive and interpret visualized data, she aims to improve visualization design, data storytelling, and data-driven decision-making.

**Marti A. Hearst** is a Professor and interim Dean at UC Berkeley School of Information, as well as a professor in the Computer Science Division at UCB. She primarily conducts research at the intersection of language and visualization in addition to work on scholarly document reading and search.